\documentclass[]{spie}  

\usepackage{amsmath,amsfonts,amssymb}
\usepackage{graphicx}
\usepackage[colorlinks=true, allcolors=blue]{hyperref}
\usepackage{comment}
\usepackage{multirow}
\usepackage{booktabs}

\title{Update on the Preliminary Design of SCALES: the Santa Cruz Array of Lenslets for Exoplanet Spectroscopy}

\author[a,b]{R. Deno Stelter}
\author[a,b]{Andrew J. Skemer}
\author[c]{Steph Sallum}
\author[a]{Renate Kupke}
\author[a]{Phil Hinz}
\author[d]{Dimitri Mawet}
\author[a,b]{Rebecca Jensen-Clem}
\author[a]{Christopher Ratliffe}
\author[a]{Nicholas MacDonald}
\author[a]{William Deich}
\author[a]{Gabriel Kruglikov}
\author[e]{Marc Kassis}
\author[e]{Jim Lyke}
\author[b]{Zackery Briesemeister}
\author[b]{Brittany Miles}
\author[a,b]{Benjamin Gerard}
\author[f]{Michael Fitzgerald}
\author[g]{Timothy Brandt} 
\author[h,i]{Christian Marois}

\affil[a]{UC Observatory, UC Santa Cruz, 1156 High Street, Santa Cruz, CA 95112 USA}
\affil[b]{Dept of Astronomy, UC Santa Cruz, 1156 High Street, Santa Cruz, CA 95112 USA}
\affil[c]{Dept of Physics and Astronomy, UC Irvine, 4129 Frederick Reines Hall, Irvine, CA 92697 USA}
\affil[d]{Dept of Astronomy, California Institute of Technology, 1200 E. California Blvd. Pasadena, CA 91125}
\affil[e]{W. M. Keck Observatory, 65-1120 Mamalahoa Hwy., Kamuela, HI 96743}
\affil[f]{Dept of Physics and Astronomy, UCLA, 475 Portola Plaza, Los Angeles, CA USA 90095}
\affil[g]{Dept of Physics, UC Santa Barbara, Santa Barbara, CA USA 93106}
\affil[h]{NRC-Herzberg Astronomy \& Astrophysics, 5071 West Saanich Rd, Victoria, BC V9E 2E7, Canada}
\affil[i]{University of Victoria, 3800 Finnerty Rd, Victoria, BC V8P 5C2, Canada}

\authorinfo{Further author information: (Send correspondence to Deno Stelter)\\E-mail: deno@ucolick.org}

\begin{document} 
\maketitle

\begin{abstract}
SCALES (Santa Cruz Array of Lenslets for Exoplanet Spectroscopy) is a 2-5 micron high-contrast lenslet integral-field spectrograph (IFS) driven by exoplanet characterization science requirements and will operate at W. M. Keck Observatory. 
Its fully cryogenic optical train uses a custom silicon lenslet array, selectable coronagraphs, and dispersive prisms to carry out integral field spectroscopy over a 2.2 arcsec field of view at Keck with low ($<300$) spectral resolution. 
A small, dedicated section of the lenslet array feeds an image slicer module that allows for medium spectral resolution ($5000-10 000$), which has not been available at the diffraction limit with a coronagraphic instrument before. 
Unlike previous IFS exoplanet instruments, SCALES is capable of characterizing cold exoplanet and brown dwarf atmospheres ($<600$ K) at bandpasses where these bodies emit most of their radiation while capturing relevant molecular spectral features.
\end{abstract}


\section{INTRODUCTION}
\label{sec:intro} 
\begin{figure}[hp]
    \centering
    \includegraphics[width=0.98\textwidth]{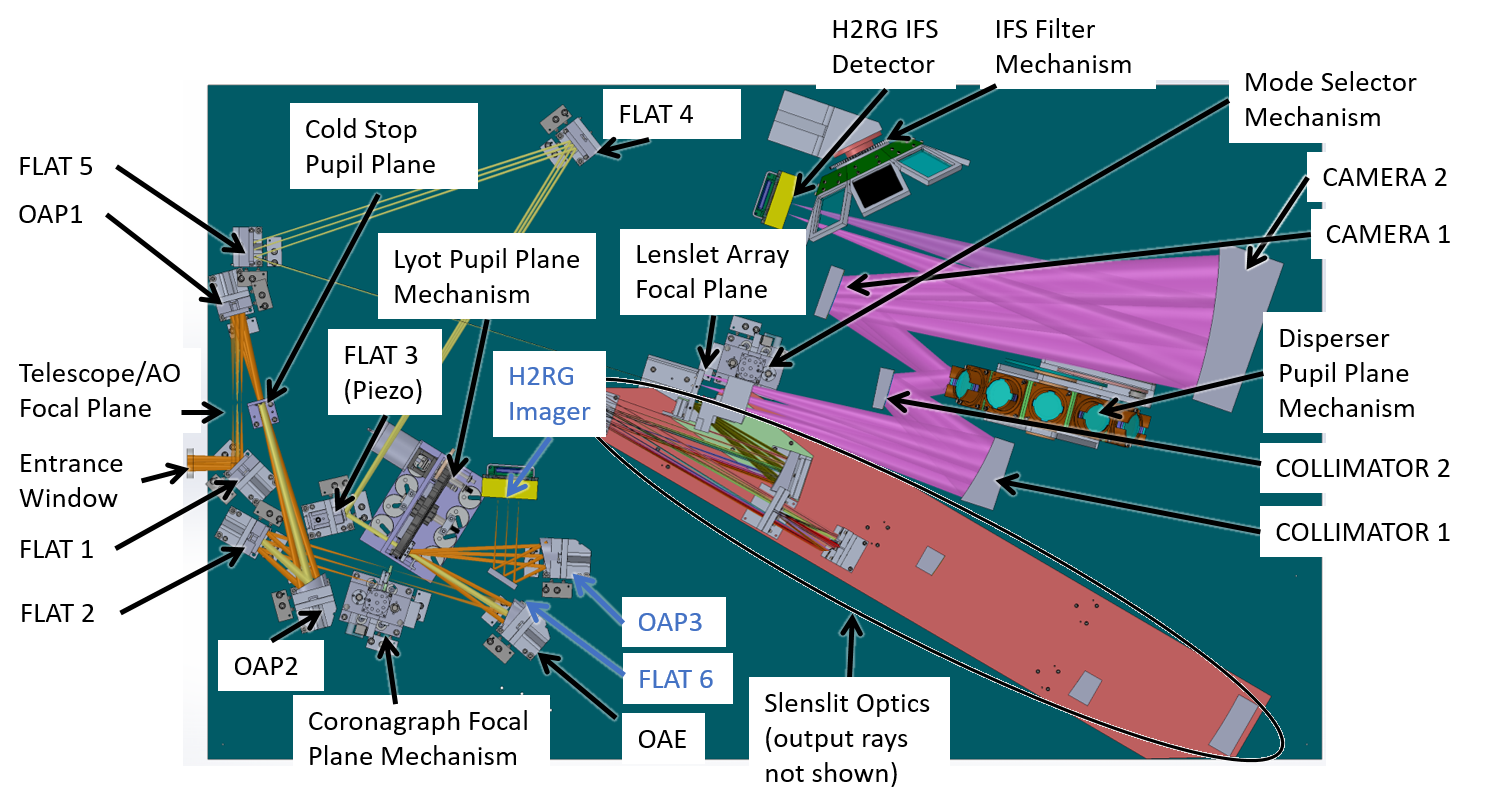}
    \caption{Annotated CAD model of the cold volume of SCALES.
        The cryostat, cold heads, and thermal shields are not shown.
        The optical bench dimensions are 1.15 m by 1.9 m and is colored green.
        Optics and cryo-mechanisms are labeled.
        The imaging upgrade optics are labeled in blue.
        }
    \label{fig:annotatedCAD}
\end{figure}

SCALES\footnote{PI: Andrew Skemer; Instrument Scientist: R. Deno Stelter} is a purpose-built exoplanet characterizer designed to sit behind either of the 10-meter Keck Adaptive Optics (AO) systems and is currently in the Preliminary Design phase (Figure~\ref{fig:annotatedCAD} shows the current CAD design). 
The high-level specifications are found in Table~\ref{tab:scalesSpecs}. 
When deployed to Keck, it will image and disperse a $2.2$ arcsec field of view at the diffraction limit from $2-5 \mu$m using a custom silicon lenslet array, selectable coronagraphs, and dispersive prisms at low ($<300$) spectral resolution.
The optical train, including the mechanisms, are fully cryogenic, with the last warm optic in the optical train being the entrance window.
A medium spectral resolution mode, with R $\approx 5000-10000$, is made possible by reimaging a subset of lenslet pupil images subtending $0.3$ arcsec by $0.3$ arcsec onto an advanced image slicer\cite{}, which reformats the square subset into a pseudo-slit; the same spectrograph optics are used between the low- and mid-resolution modes, with the prisms being swapped out for gratings.
\begin{table}[hp]
    \centering
    \caption{The high-level specifications of SCALES.}
    \includegraphics[width=0.9\textwidth]{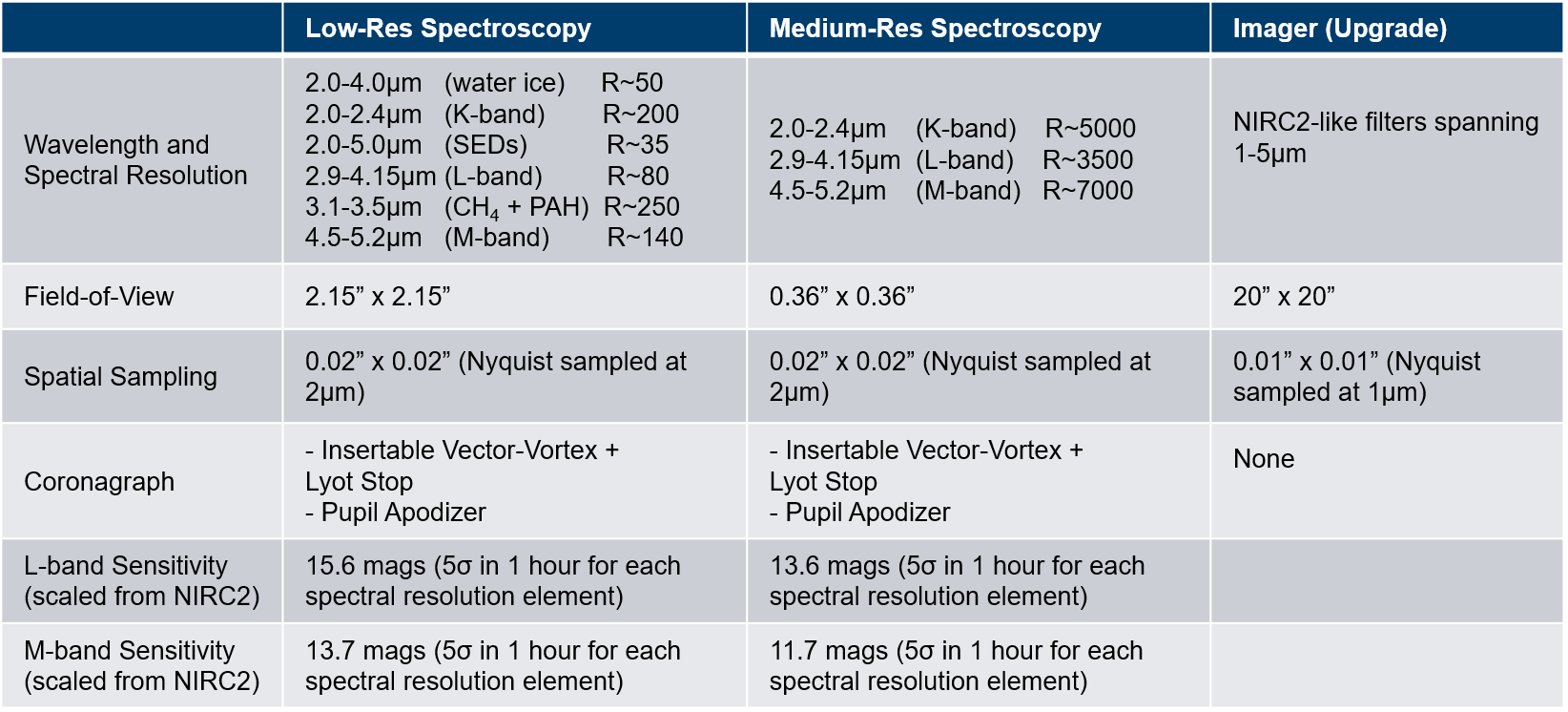}
    \label{tab:scalesSpecs}
\end{table}

An imaging arm `piggybacks' off of first portion of the optical train and feeds a dedicated detector with a $20$ arcsec by $20$ arcsec field of view with diffraction-limited performance from $1-5 \mu$m. 
Both the spectrograph and the imaging arm will use Teledyne HAWAII-2RGs (H2RG) with a $5.3 \mu$m cut-off as the detectors.
The readout electronics are Teledyne SIDECAR digitizers, which are controlled with MACIEs mounted to the exterior to the vacuum vessel.
UCLA is developing the detector head packaging based on the SCAM detector~\cite{martin2018scam}.

SCALES is similar to the conceptual TMT instrument Planetary System Imager's red arm (PSI-Red~\cite{PSIbase2018,PSIredAndy2018}) and in principle, could be upgraded and moved to TMT.
In this proceeding we discuss science drivers (\S\ref{sec:scienceDrivers}) and the design of both the low and medium spectral modes and the imager in \S\ref{sec:optdesign} in more detail.
Z. Briesemeister's paper (this session, \#11447-294) describes the end-to-end simulation of science targets with SCALES.
The opto-mechanical design of each module, along with the cryo-mechanisms and cryostat, is discussed in \S\ref{sec:optmechdesign}.

\section{SCIENCE DRIVERS}
\label{sec:scienceDrivers}
\subsection{SCALES’s Exoplanet Imaging Advantage}
SCALES combines the two most powerful methods for imaging exoplanets: thermal infrared ($2-5 \mu$m) imaging, which detects exoplanets at wavelengths where they are bright~\cite{skemer2014}, and integral-field spectroscopy, which distinguishes exoplanets from residual starlight based on the shapes of their spectral energy distributions. 
For a 300~K planet, this combination creates a $\sim4-5$ magnitude boost in sensitivity compared to an H-band IFS, and a $\sim2.2$ magnitude boost in sensitivity compared to an L-band imager.\footnote{
For a given planet temperature, we estimate planet colors in common bandpasses (imaging) or optimal bandpasses (IFS), using models of Y dwarfs and exoplanet atmospheres~\cite{Morley2014}, and calculate contrast with respect to a Raleigh-Jeans tail. 
The optimal filter provides an effective contrast-boost for integral field spectroscopy~\cite{Skemer2018a}. 
Additionally, for integral field spectroscopy, we adopt a 1.0 magnitude gain that roughly encompasses the empirically demonstrated bonus of IFS-based star-light speckle suppression techniques~\cite{Marois2014,Vigan2015,wahhaj2018}, which are independent of optimal filtering.}
By operating at longer wavelengths than other high-contrast integral-field spectrographs (e.g. GPI~\cite{macintosh2014gpi}, SPHERE~\cite{beuzit2008sphere}, and CHARIS~\cite{groff2015charis};  SCALES will extend the wavelength range we use to characterize planets, and also discover new planets (in particular, cold planets) that are not detectable with near-infrared instruments. 
Despite the competitiveness of the exoplanet imaging field, SCALES’s unique parameter space ensures that it will lead a broad range of new science.

\subsection{Exoplanet Imaging Surveys with SCALES}
SCALES will detect and characterize exoplanets in several categories, including some new categories that have not been accessible to imaging before (e.g. cold planets, and planets with well-constrained masses):
\vspace{-0.5mm}
\begin{itemize}
    \item \textbf{Gaia Exoplanets and Y-Dwarfs} -- 
        Around the time SCALES is completed, Gaia’s extended 9-year survey will be releasing a catalog of ~70,000 exoplanets~\cite{Perryman2014gaia} 
        Some of these planets will be around nearby stars at wide enough separations to be directly imaged.  
        Using the SCALES predicted contrast curve (based on the NIRC2 contrast curve~\cite{mawat2019}), and GAIA data~\cite{brandt2019astronometry}, we predict 9 planets ($<13 M_{jup}$) and 67 brown dwarfs ($>13 M_{jup}$) will be detectable by SCALES after GAIA’s 9-year survey.  
        With an additional epoch from WFIRST in 2030, this increases to 19 planets and 144 brown dwarfs.  
        This is particularly exciting, since in general, direct imaging cannot measure planet/brown dwarf masses, and atmospheric properties are degenerate without known masses. 
        Planet orbits measured via imaging can also lead to stellar mass estimates~\cite{Nielsen2014}.
        Many of the new Gaia planets and brown dwarfs amenable to direct imaging will be $\sim300$K (cold enough for water clouds to be visible in M-band spectra~\cite{skemer2016coldestbrowndwarfspectrum}.
        These cannot be imaged with today’s near-infrared high-contrast IFSs, but can be imaged at longer wavelengths~\cite{skemer2014}.
\vspace{0mm}
    \item \textbf{Young ($\mathbf{<100}$ Myr), Cold ($\mathbf{<600}$K) Exoplanets} -- 
        The most common way to search for directly-imaged planets is to look around young stars, where exoplanets are thought to be relatively warm before they radiate away the residual heat from their formation~\cite{Biller2007, vigan2012}.
        However, planets that form by core accretion may radiate away their energy quickly through accretion shocks, and thus form with relatively cold initial conditions ($<600$K).  
        As a result, there may well be a large population of ``cold-start'' planets lurking just below the sensitivity of near-infrared IFS’s~\cite{stone2018LEECH}, which have only imaged planets as cold as $\sim600-750$K~\cite{Macintosh2015}. 
        SCALES will be sensitive to this hypothetical population, which would complement the young, warm planets discovered by previous surveys, and the old, cold planets that will be discovered by Gaia.
\vspace{-2.5mm}
    \item \textbf{Accreting Protoplanets} -- 
        By observing young planets that are still embedded in their nascent disks, it is possible to observe mass accretion directly. 
        To date, only a couple of accreting exoplanets and candidates have been discovered~\cite{Kraus2012LkCa,Keppler2018}.
        Part of the difficulty of imaging accreting protoplanets is that they are usually embedded in protoplanetary disks, whose scattered light can resemble exoplanets~\cite{Sallum2016,Currie2019}.
        Integral-field spectroscopy, specifically at the wavelengths where these planets peak in brightness ($4 \mu$m~\cite{Sallum2015}), will allow us to distinguish between circumstellar disk scattered light and protoplanet emission (ongoing AO upgrades at Keck are also geared towards imaging planets around young stars~\cite{bond2018irpyramid}).
\vspace{-2.5mm}
    \item \textbf{Known Exoplanets} -- 
        A recent review lists 47 directly-imaged planetary-mass or near-planetary-mass companions~\cite{bowler2016} and there are a similar number of more massive brown dwarf companions.  
        SCALES will extend spectroscopy of these planets into the thermal infrared, enabling accurate bolometric luminosity measurements, atomic abundance measurements that span all relevant species, and a broad-wavelength lever-arm for measuring the slowly varying opacities of clouds.
\end{itemize}

\subsection{Exoplanet Characterization with SCALES}
To study exoplanets beyond their locations, masses and radii requires photometry and spectroscopy of the planets themselves. 
Spectroscopy can reveal an exoplanet’s thermal structure, chemical composition, cloud properties, spatial inhomogeneity, and much more~\cite{barman2011,marley2012}.
The observational signatures of these properties are often degenerate and require broad wavelength coverage to disentangle~\cite{skemer2014,skemer2016LEECH}.
Clouds, which are ubiquitous on exoplanets~\cite{marley2013hazes} have optical properties that vary slowly with wavelength, and can be confused with thermal structure effects over a narrow bandpass~\cite{line2016}.
Molecules can probe chemical reactions, vertical mixing and atomic abundances such as C/O ratios~\cite{konopacky2013}, but only if a relatively complete set are measured over the broad wavelength range where they are individually detectable~\cite{barman2015}.

SCALES will enable these sorts of broad wavelength studies by complementing existing near-infrared high-contrast IFS’s (e.g. GPI, SPHERE, and CHARIS) as well as some lower contrast near-infrared IFS’s that have higher spectral resolution (OSIRIS~\cite{larkin2006}, SINFONI~\cite{eisen2003sinfoni}. 
Medium-resolution spectroscopy is a new high-contrast capability that SCALES will provide for the first time at any wavelength, allowing line-by-line identification of molecules, while simultaneously providing broad bandpass continuum measurements, all within the environment of a coronagraphic IFS.

\subsection{Non-Exoplanet Science}
Thermal infrared imaging and spectroscopy are used for a wide range of Solar System, Galactic, and extragalactic observations. 
SCALES integral-field spectroscopy will provide a novel and powerful extension. 
Because thermal infrared observations are in almost all cases background limited, there is no signal-to-noise penalty for dispersing images into spectroscopic data cubes. 
For observations where SCALES’s field-of-view is big enough, IFS data-cubes provide gains over straight imaging with essentially no downside. 
Keck’s current thermal infrared imager, NIRC2, publishes $\sim70$ papers per year, the vast majority of which entail adaptive optics imaging of structures that are smaller than the SCALES field-of-view. 
We list examples of topics that will benefit from thermal infrared integral-field spectroscopy: 

\begin{itemize}
\vspace{-4mm}
    \item \textbf{Volcanic Eruptions on Io} -- 
        SCALES will monitor the locations, temperatures and areal extents of volcanoes on Io, which emit at $>3 \mu$m. For large eruptions, SCALES will provide spatially resolved temperature maps.
\vspace{-3mm}
    \item \textbf{Ice Lines in Disks} --
        SCALES will observe protoplanetary and debris disks to measure PAHs ($3.3 \mu$m) and water ice ($3.1 \mu$m), a primary building block of planetary cores.
        Infrared water ice measurements will complement ongoing ALMA water vapor disk observations~\cite{podio2013}.
\vspace{-3mm}
    \item \textbf{Thermal lines and PAHs in Supernovae} -- 
        SCALES will observe Brackett, Pfund and CO-lines ($2-5 \mu$m) in bright young supernovae. 
        SCALES will observe carbonaceous dust in the unshocked ejecta of nearby remnants to study dust formation and destruction in supernovae.
\vspace{-3mm}
    \item \textbf{Spatially Resolved Spectra of Nearby Galaxies} -- 
        SCALES will spatially resolve the nuclear regions and star-forming regions of nearby bright galaxies, AGN, ULIRGs, etc., probing PAH features and hot dust emission as a window on dust heating and small dust grain production/destruction.
\end{itemize}

\section{OPTICAL DESIGN}
\label{sec:optdesign}
SCALES is designed to image exoplanets using tried-and-tested optical and opto-mechanical design principles based around an all-reflective design (excepting, of course, the entrance window, coronagraph masks, lenslet array, dispersers, and filters, which must be refractive).
The optical train of SCALES is effectively 3 collimator-camera relays in sequence, with two collimator-cameras in the fore-optics (where the coronagraph and apodizing Lyot stop are, see \S\ref{subsec:foreoptics}) that illuminate the lenslet array (see \S\ref{subsec:lensletArrayOpticalDesign}), followed by the spectrograph module (see  \S\ref{subsec:spectrograph}) which is the third collimator-camera.
We have designed SCALES in Zemax in two portions: the fore-optics and the spectrograph modules.
Future work will develop a combined model using a combination of sequential and nonsequential models.

The spectrograph module is designed to match the fast beam output of the lenslet array with excellent image quality preserved across the field of view.
We have also designed a module (dubbed a slenslit, see \S\ref{subsec:slenslit}) that relays a small subset of lenslet array pupil images into a psuedo-slit, which feeds the spectrograph module with a pseudo-slit that we disperse at higher spectral resolution.
Because the optical train of the fore-optics is all-reflective, we were able to design a high-performance imaging mode as a possible upgrade with a field of view of $20$ arcsec on a side by increasing the size of the mirrors of the fore-optics and only adding in one additional powered optic; this upgrade is discussed in \S\ref{subsec:imager}. 

SCALES is designed with modular upgrades in mind.
We are pursuing the possibility of a Self-Coherent Camera \cite{Gerard2018} which would allow for in-instrument wavefront sensing as an upgrade and is described in \S\ref{subsec:scc}.
This would allow us to effectively remove the non-common path aberrations (NCPA) by having wavefront sensing closer to the sampled focal plane than otherwise.
Another upgrade is to add in a single-mode (SM) fiber bundle that feeds a high-resolution (R$\sim100000$) spectrograph (see \S\ref{subsec:hchd}, high contrast high dispersion spectroscopy~\cite{hchd2017}).
This would require an actuated pick-off mirror and reimaging optics, similar to the KPIC design~\cite{kpic2018,Pezzato2019}, and feed a separate spectrograph.
The opto-mechanical design is discussed in \S\ref{sec:optmechdesign}.

\subsection{Fore-Optics}
\label{subsec:foreoptics}
\begin{figure}
    \centering
    \includegraphics[width=0.8\textwidth]{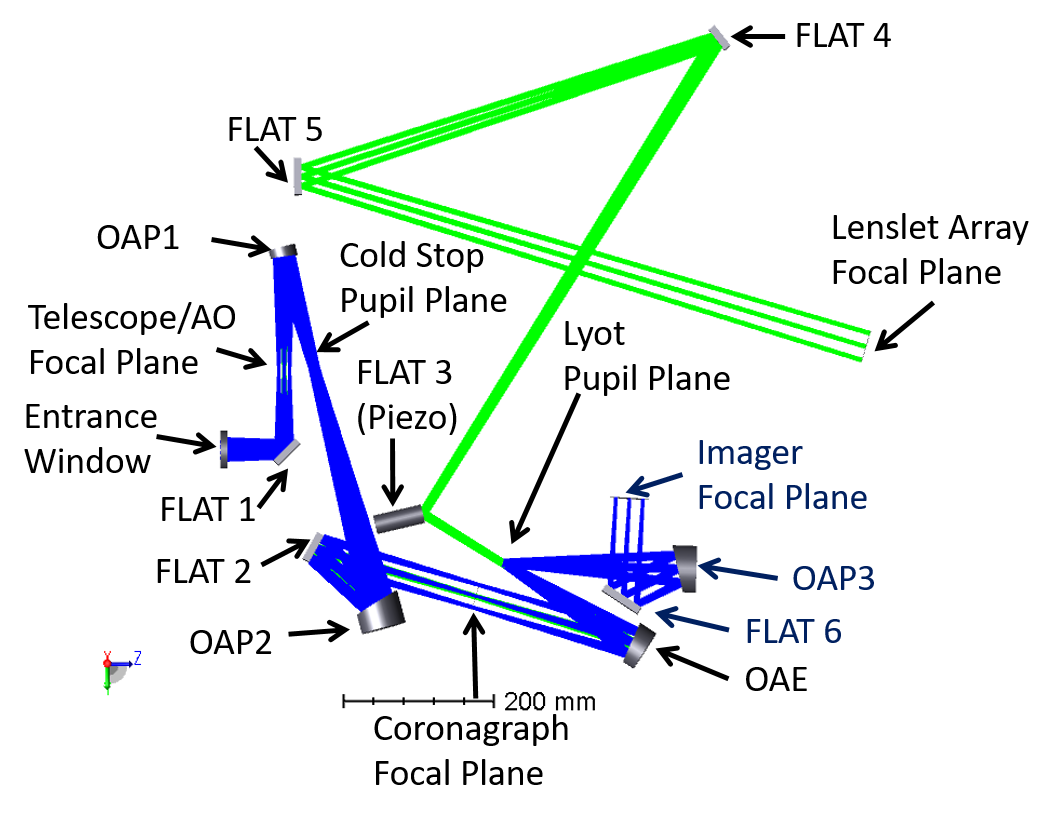}
    \caption{The SCALES fore-optics design from Zemax.
    Light enters from the left through the entrance window and propagates through the optical train.
    Note that the fore-optics that are in common (all of them up to the Lyot pupil plane) are all oversized relative to the IFU beam in order to accommodate the imager beam footprints.
    The lenslet array is the final optic and lives at the focal plane at the right of the image.}
    \label{fig:foreoptics}
\end{figure}

We began the optical design using the as-built optical design of the Keck telescope and AO bench.
Figure~\ref{fig:foreoptics} shows the Zemax model, with the IFU beams shown in green and the imager shown with blue beams.
The optical design of the SCALES fore-optics requires at minimum a focal and pupil plane suitable for coronagraphic masks.
This, in turn, requires both coronagraphic mask planes being placed behind a cold stop to limit background radiation from sources outside the telescope beam (e.g., the telescope structure, AO bench and warm optics thereon, telescope dome, atmosphere, etc).
As a result, the AO-provided focal plane is re-imaged by an OAP onto the cold stop.

This pupil plane is reimaged by another OAP onto an f/26 focal plane, which hosts the focal plane coronagraph masks.
The masks are carried by a linear slide described later in \S\ref{subsec:cryomechs}.
There are 5 mask positions (one each for K, L, M,  a lens for pupil imaging, and a mask position to be used with experimental coronagraph designs).

After the second focal plane is an Off-Axis Ellipse (OAE) mirror, which provides both a second pupil plane (where we place our pupil coronagraphic masks) and the f/350 focal plane which hosts the lenslet array; this OAE removes the need to use two OAPs, simplifying the optical design.
The second pupil plane is populated with a rotary mechanism (the `Lyot wheel') that holds up to 15 pupil masks to be used in conjunction with the focal plane masks.

A total of five flat fold mirrors are included in the IFU path (the first two are in common with the imager arm) and are included to better package the optical train.
We have completed a preliminary tolerancing analysis of the optics and find that all mirrors are well within nominal specifications of manufacturers.
The RMS wavefront error (WFE) at the lenslet array is
\begin{figure}
    \centering
    \includegraphics[width=0.98
    \textwidth]{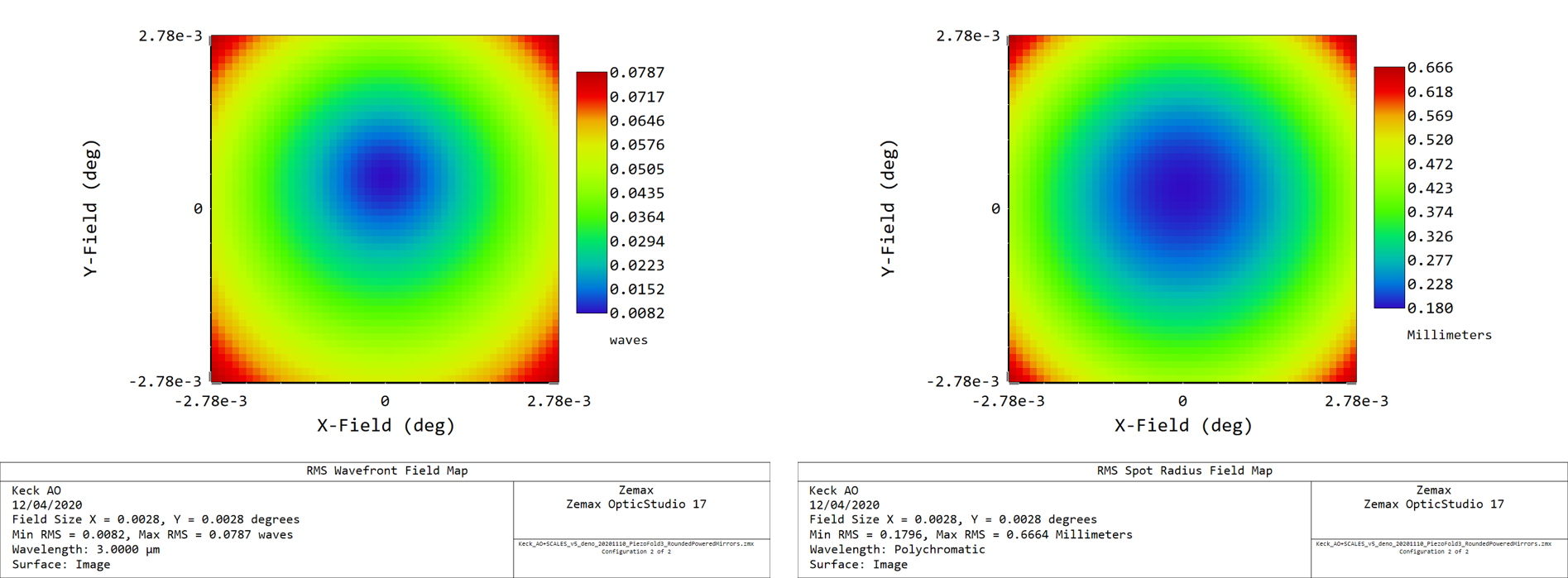}
    \caption{Optical performance of the fore-optics at $3 \mu$m at the lenslet array focal plane.
    Left: RMS Wavefront Error map.
    The nominal RMS WFE is worst at the corners (shaded red) but remains $<0.1$ waves and averages $<0.05$ waves across the field of view.
    Right: RMS Spot size.
    The lenslets are sized to Nyquist-sample the field of view at $2 \mu$m ($\sim340\mu$m in size).}
    \label{fig:rmswfemap}
\end{figure}
A more detailed tolerancing analysis will inform us of the tolerances of the mid-spatial frequency errors which arise from resonances that form during the diamond-turning process, but we expect that they will be manageable as we have left ample room in our error budget above and beyond the limits imposed on the initial tolerancing analysis.

The high magnification (roughly a factor of 23.5) is required by the desire to have a 2.2 arcsec field of view fill an H2RG; the spectrograph module is designed  to have a magnification of unity.
The footprint of the beam on the lenslet array, therefore, is the same size as an H2RG-18 (a square 36.72 mm on a side, taking into account the 4 pixel-deep reference picture frame).
The lenslets are designed to be squares $341 \mu$m on a side with an f/8 beam at the output.
The lenslet array outputs a regular grid of pupil images, one per lenslet, which are then dispersed onto the detector (with a prism or grating).
By rotating the prism about the optical axis, we can control the angle at which the spectra are dispersed.
Because each lenslet produces a spectrum, and because we can assign a set of spatial coordinates to each lenslet, each spectrum can be thought of as a spaxel, or spectral pixel.

In order to accommodate the slenslit (discussed further in \S\ref{subsec:slenslit}), we have designated fold mirror 3 to be powered by a piezoelectric tip/tilt stage (note that it is shown as an elongated cylinder in Figure~\ref{fig:foreoptics}).
This tip/tilt stage, being in collimated space close to the pupil, steers the focal plane about on the lenslet array and allows us to send exoplanetary light through the slenslit to be dispersed at much higher spectral resolution than allowed by the low-resolution mode.

\subsection{Lenslet Array}
\label{subsec:lensletArrayOpticalDesign}
\begin{figure}
    \centering
    \includegraphics[width=.98\textwidth]{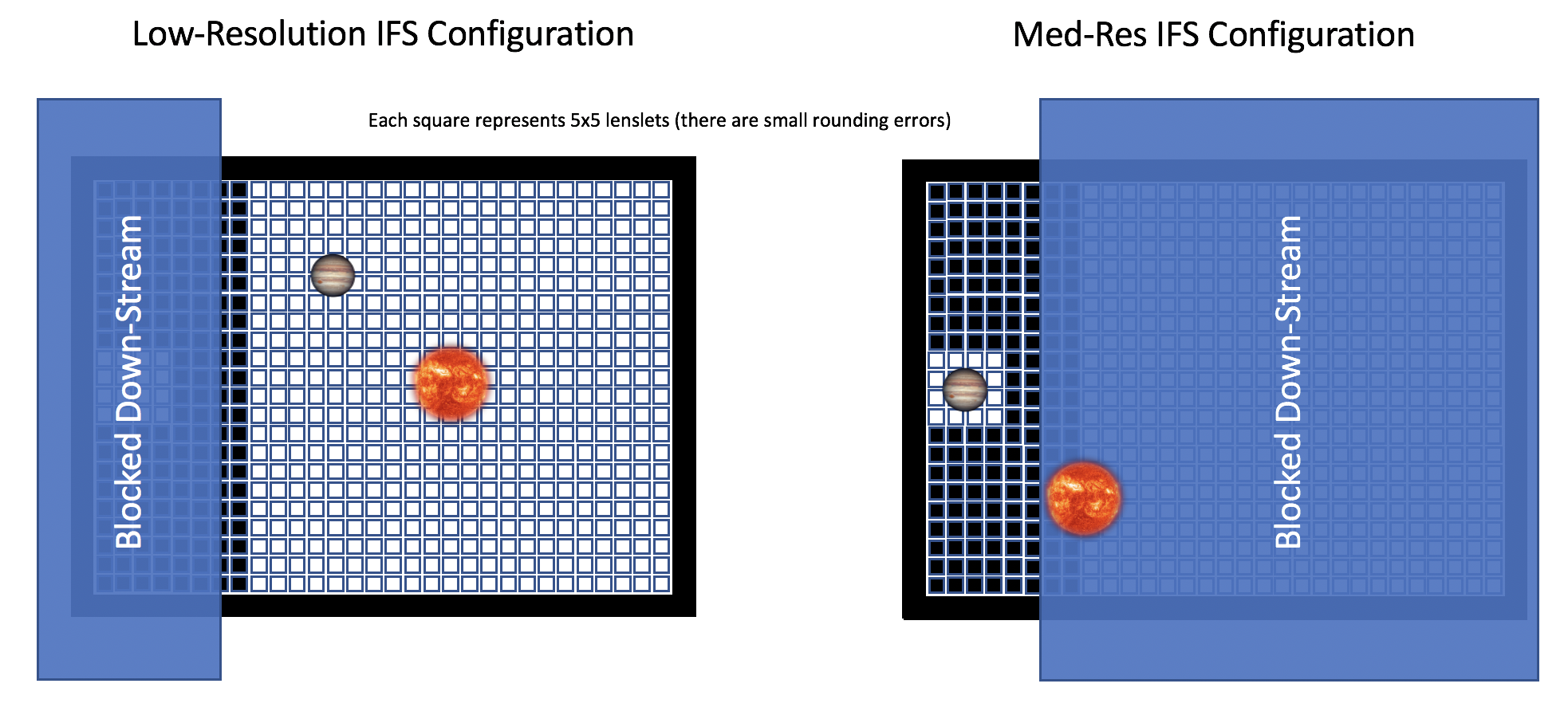}
    \caption{Schematic showing the lenslet array being used in its low-resolution configuration and its medium resolution configuration.  
    Left: In the low-resolution configuration, the piezo tip-tilt stage steers the star (which has already been suppressed by the coronagraph) to the middle of the larger subarray and the exoplanet rotates around the star with parallactic angle.  
    Right: In the medium-resolution configuration, the piezo steers the exoplanet onto the smaller subarray and holds it there with small tip-tilt adjustments as the exoplanet rotates around the star.}
    \label{fig:lensletschematic}
\end{figure}

The SCALES lenslet array contains two regions: a 110x110 lenslet subarray that feeds the low-resolution spectrograph and an 18x18 lenslet subarray that is picked -off by the medium resolution slenslit optics, with the remaining lenslets painted black (See Figure \ref{fig:lensletschematic}). 
Both configurations fully illuminate an H2RG detector with spectra.  
On the fore-optics side, a piezo steers the intended field onto either the low-resolution subarray or the medium-resolution subarray.  
On the spectrograph side, a configuration-selector transmits either the low-resolution or medium-resolution beams.  
Behind each lenslet is a pinhole that truncates the square diffraction spikes of each lenslet~\cite{Peters2012charis}, which is particularly important given SCALES's relatively long wavelength.  
The lenslet array and pinhole array are each photolithographically etched silicon substrates, which are aligned and glued to a silicon picture frame, as was done with the precursor instrument, ALES~\cite{skemer2015}.

The physical sizes of the lenslets is determined by the desired length and separation of spectra on the H2RG~\cite{bacon1995}.
Following the equations described in the ALES paper~\cite{skemer2015}, we select a spectrum length of 60 pixels for the  low-resolution mode, with each spectrum separated by 6 pixels, corresponding to a lenslet size of $341\times341\mu$m.  
With an f/8 spectrograph and pinholes, this separation eliminates most crosstalk between neighboring spaxels, which is important for high-contrast imaging.  
The medium resolution mode uses the same size lenslets and interleaves them to create spectra that have a length of $\sim$2000 pixels that are, again, separated by 6 pixels.
The physical size of the lenslets, along with the requirement that SCALES Nyquist sample its shortest wavelength (2.0 $\mu$m for the IFS) sets the magnification in the fore-optics.  

\subsection{Spectrograph}
\label{subsec:spectrograph}
The spectrograph is a telecentric one-to-one reimaging collimator-camera pair operating at f/8 (matching the lenslet array output).
The lenslet array pupil images at the pinhole grid are reimaged onto the detector.
Both collimator and camera are made up of two freeform surfaces for four mirrors total; there are no fold mirrors.
\begin{figure}
    \centering
    \includegraphics[width=.98\textwidth]{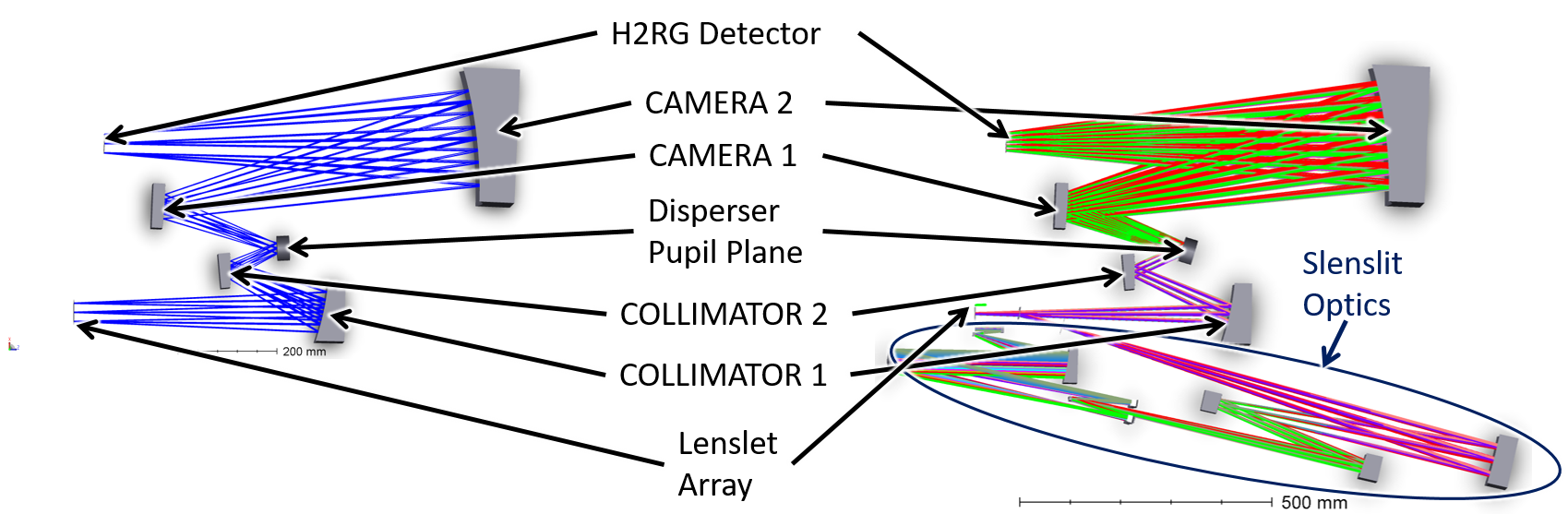}
    \caption{The SCALES spectrograph optical design from Zemax.
    Left: The low-resolution spectrograph optical layout.
    The lenslet array is at the lower left, with light propagating to the right.
    Right: The slenslit optical design described in \S\ref{subsec:slenslit}.
    The spectrograph collimator and camera optics are the same as for the low-resolution mode, but light is relayed through the slenslet optics before passing to the spectrograph optics.}
    \label{fig:spectrograph}
\end{figure}
The spectrograph collimator produces an unobscured pupil image with a geometric semi diameter of $\sim25$ mm.
A selectable carousel holds up to 10 dispersers (prisms, grisms, or gratings), an optical flat mirror, and an engineered diffuser for a total of 12 positions.
This pupil plane is reimaged by the camera onto the H2RG spectrograph detector.
A selectable filter mechanism sits $40$ mm in front of the detector and hosts bandpass selector filters.
Performance across the field of view is diffraction-limited in both the low-resolution and mid-resolution mode.

The dispersers for the low-resolution mode will be LiF prisms.
LiF was chosen for its excellent transmission and reasonably linear dispersion across the $2-5 \mu$m bandpass.
The prisms are designed to be double-pass; the back surface will be gold-coated while the front surface will have a custom AR coating unique to each prism.
The mid-resolution mode requires using custom 1st-order gratings, which will be mounted to the carousel mechanism.

\subsection{Slenslit}
\label{subsec:slenslit}
Once the field has been sampled at the lenslet array, image quality of the optics can be loosened for the spectrograph optics.  Spectral resolution is then limited by the pitch of the lenslets which limits the trace length between spectra before they overlap at the detector.
One could select only single columns (or rows) to produce a pseudo-slit, effectively turning the lenslet array into a longslit spectrograph at the cost of the field of view and decreased signal to noise (given that lenslet arrays are in general are Nyquist-sampled).
While going to a higher natural number~\cite{skemer2015} is an option,  one cannot spread the light from exoplanets across the entire detector without having neighboring spaxels overlap using traditional lenslet IFU designs.
If one were free to rearrange the locations of the lenslets such that their dispersed spectra weren't overlapping with neighboring lenslets' spectra, we could push to much higher spectral resolution.
This is where the slenslit (\textbf{S}liced \textbf{LENS}let pseudo-s\textbf{LIT}) idea was born.

One of the other techniques for building IFUs is to use slicing optics.
Advanced image slicers \cite{fisica2004,FRIDA2013,miradas2016} use powered optics machined out of monolithic blocks of metal to geometrically rearrange a 2-dimensional field of view into a linear pseudo-slit, which is then be dispersed by a spectrograph.
The lenslet IFU has significantly looser specifications on the spectrograph optics because a lenslet array destroys the spatial information otherwise preserved by the slicing optics, which would otherwise be forced by the spectrograph optics to preserve all the way to the detector.

In the case of SCALES, we have combined the two IFU techniques to provide a spectral resolution of R$\sim5000 - 10000$ using an all-spherical, monolithic advanced image slicer.
We illuminate an 18x18 patch of lenslets with exoplanetary light (selected by actuating  the third fold mirror close to the second pupil plane with a piezoelectric tip/tilt stage) and re-image the lenslet pupil images onto an 18-slice slicer mirror array using a Three Mirror Anastigmat (TMA) relay.\footnote{
The relay TMAs have aspherical elements but are designed to be relatively simple to fabricate.}
\begin{figure}
    \centering
    \includegraphics[width=0.95\textwidth]{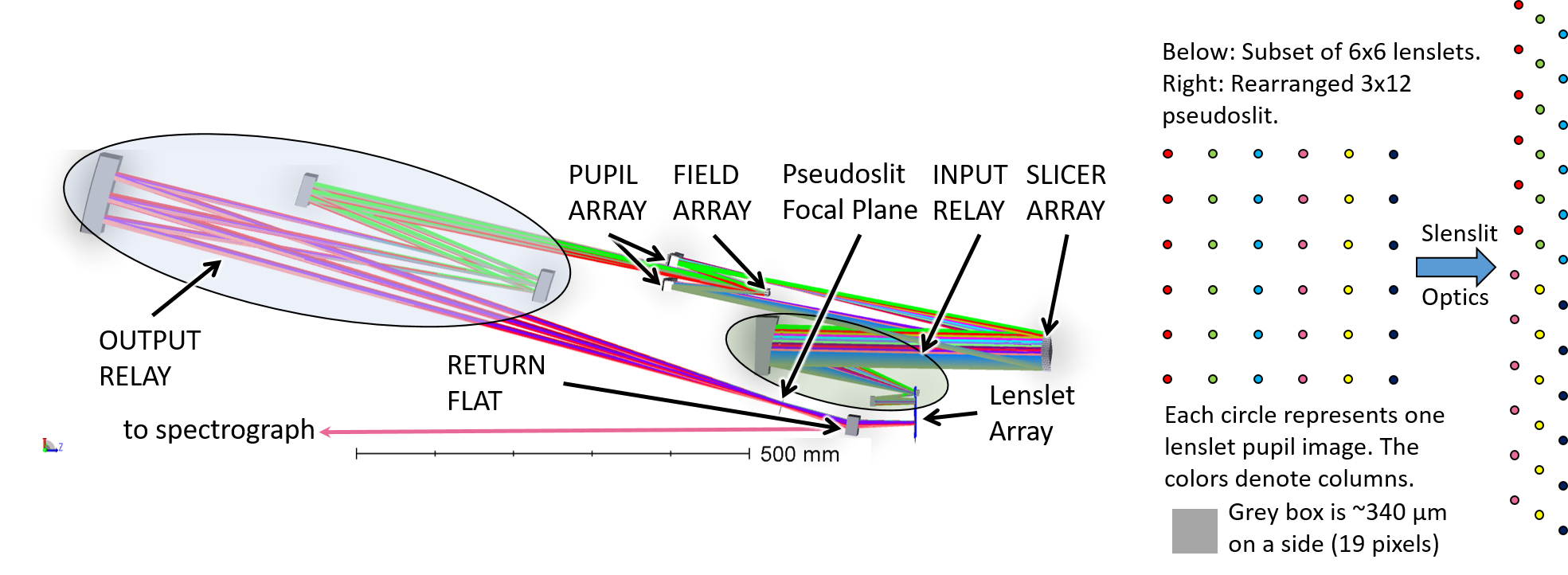}
    \caption{The SCALES slenslit.
    Left: Annotated optical design.
    Note the orientation is flipped relative to Figure~\ref{fig:spectrograph}.
    Right: Schematic of the slenslit input and output. 
    A $6\times6$ subset of lenslets is geometrically rearranged via the slenslit optics (denoted by the blue arrow) into a $3\times12$ pseudoslit.
    The SCALES slenslit will rearrange an $18\times18$ subset of lenslets into a $3\times72$ pseudoslit.
    The `packing density' in the vertical direction of the pseudoslit is increased by a factor of 3 relative to the unaltered lenslet array by staggering the lenslet columns vertically in the pseudoslit.}
    \label{fig:slenslitODdesign}
\end{figure}
This relay allows the slicer to operate at f/64 (as image slicers in general have less aberration at slower beam speeds). 
2 columns of 9 pupil mirrors each reimage the 18 columns of lenslets onto a field mirror array consisting of 3 columns of 6 rows of mirrors while also decreasing the beam speed from f/64 to f/32.\footnote{
The pupil mirrors are at the pupil plane produced by the slicer mirrors, while the field mirrors are at the focal plane produced by the pupil mirrors.}
At the field mirrors the lenslet columns are slightly staggered in the vertical direction such that the vertical spacing between lenslets is reduced from $341 \mu$m to $110 \mu$m, which is about the `natural' spacing of the spaxels of the low-resolution mode.
Thus, the slicing optics increase the fill density of the lenslets by a factor of 3, making a pseudo-slit suitable for much higher dispersion than offered by the classic lenslet IFU low-resolution mode over a (small) two-dimensional field of view.
The output relay reimages light from the field mirrors to a return fold mirror on a linear stage (described in \S\ref{subsubsec:modeSelectorSlide}) further increasing the beam speed from f/32 to f/8 to match the spectrograph's expected input.
The spectrograph optics then reimage and disperse the pseudoslit onto the detector by using custom gratings.

Each slice (or column of lenslets)  is designed and optimized separately before being combined into the overall slenslit design.
Verification of the performance required adding the slenslit optics to the spectrograph optical design, whereupon another round of slight tweaks and optimization to ensure all slices share a common pupil at the disperser was required.
The performance is diffraction-limited across the field of view of 18x18 lenslets and allows us to go to much higher spectral resolution never before reached in high contrast IFU spectroscopy.
The optical design of of the slenslit is shown in Figure~\ref{fig:spectrograph} and opto-mechanical design in Figure~\ref{fig:slenslitODdesign}.

\subsection{Imaging Arm Upgrade}
\label{subsec:imager}
Given the high demands placed on the optical design of the fore-optics, adding in a diffraction-limited imager operating from $1-5 \mu$m did not affect the baseline requirements of the instrument.
In fact, very little was changed aside from the aperture size increase and the addition of three optical elements (two flats and an OAP) as shown in Figure~\ref{fig:foreoptics}.
The imaging fold is split-off at the Lyot wheel (located at the second pupil plane).  
This allows the imager to reimage the rejected stellar light for certain mask designs (specifically those that are reflective).
When SCALES is used in imaging mode without corresponding spectroscopy, the Lyot mechanism places a mirror at the pupil plane, folding the beam to the OAP and final flat fold mirror before the detector; the last fold mirror allows for better baffling and easiest packaging constraints.

The imager also requires an additional filter mechanism after the cold stop, which would be upgraded from a simple fixed cold stop to a selectable cold stop mechanism this allows for different sized cold stops to be used for imaging or low-res IFU spectroscopy).
The two mechanisms occupy the first pupil space between the first two OAP mirrors and are designed such that the filter mechanism may be removed without disturbing the cold stop; this is so that the cold stop alignment (critical for the low-resolution high contrast portion of the instrument) is not disturbed during filter changes.
The imager H2RG readout electronics are Teledyne SIDECAR ASIC boards with MACIE interfaces for commonality with the spectrograph detector package.
We are also investigating using a $10 \mu$m cutoff focal plane array in place of the $1-5\mu$m array and are testing a Teledyne device using SIDECAR/MACIE electronics at UCSC.
This benefits the SCALES project by familiarizing our team with the MACIE controller and its behavior when interfacing with the SIDECAR electronics.

\subsection{Self-Coherent Camera Upgrade}
\label{subsec:scc}
The self-coherent camera (SCC) concept \cite{scc} utilizes the same principle as the classical Young double slit experiment, interfering light between two apertures on a detector to enable sensing both the phase and amplitude of the electric field based on fringe position and intensity, respectively. 
The SCC is realized in a coronagraphic image by interfering the classical coronagraphic pupil plane (i.e., the Lyot plane) with a small off-axis pinhole aperture in the same plane, forming fringes in the coronagraphic image overlaid on the existing (unfringed) speckle pattern and thus transforming a coronagraphic imager additionally into a wavefront sensor. 
As the light transmitted through this off-axis pinhole in the Lyot plane is only from starlight (due to the inherent diffractive properties of a focal plane mask), fringes form only on coherent stellar speckles, while any incoherent exoplanet light remains unfringed, enabling both wavefront sensing and post-processing algorithms to attenuate remaining stellar speckles without affecting exoplanet throughout.

The Fast Atmospheric SCC Technique (FAST) builds off of the classical SCC design by using a custom focal plane mask to provide much higher fringe visibilities (i.e., fringed vs. unfringed image components), enabling both post-processing \cite{Gerard2018} and active wavefront sensing and control\cite{Gerard2018b, Gerard2020} down to millisecond timescales on bright stars, opening the door to correction of residual atmospheric, in addition to quasi-static, speckles. 
Several FAST focal plane masks have now been fabricated and tested in the lab \cite{Gerard2019,lardiere2020}, confirming that the simulated fringe visibilities and speckle subtraction capabilities are indeed achievable with this technique. 

We are in the initial stages of implementing FAST into SCALES, with a preliminary conceptual design planned for summer 2021 and applying for full funding (NSF ATI or similar) in fall 2021. 
FAST will likely be realized as a a narrowband H band imager, using one of the existing slots in the coronagraph slide and Lyot stop wheel and a dichroic and associated optics just downstream of the Lyot stop to relay H band light to the SCC imager. 
The OAE between the focal plane mask and Lyot stop is already sufficiently oversized to be compatible with the SCC design; no further modifications to the existing SCALES design are necessary to be compatible with an H band FAST imager. 
In addition to its wavefront sensing capabilities, FAST will also enable new H band exoplanet characterization science cases\cite{Gerard2019} that are complementary to the existing mid-infrared capabilities of SCALES.
\subsection{High Contrast High Dispersion Spectroscopy with Single-Mode Fibers}
\label{subsec:hchd}
High contrast, high dispersion spectroscopy (HCHD spectroscopy) utilizes Single-Mode (SM) optical fibers coupled to exoplanetary light after starlight suppression via coronagraphic techniques to feed a high-resolution (R$\sim100000$) spectrograph~\cite{hchd2017}.
Coupling SM fibers to exoplanetary light has been demonstrated on-sky by KPIC~\cite{Pezzato2019,kpic2018} using vortex coronagraphs and fiber nulling techniques with room-temperature optics feeding the cryogenic NIRSpec instrument.
The SCALES HCHD spectrgraph mode, in comparison to KPIC, will have higher sensitivity due to its cryogenic coronagraph and injection optics.
High-resolution SM fiber spectrographs are rather compact due to the single mode input: PARVI~\cite{PARVI2020} is $\sim0.5\times0.5\times0.5$ m\textsuperscript{3} spanning J-H (one octave), while HISPEC~\cite{hispec2019} is $\sim0.5\times0.5\times1$ m\textsuperscript{3} spanning Y-K (two octaves) and is effectively two PARVIs joined at the echelle grating.
It is likely that SCALES will opt to have the fiber coupler module inside the cryostat, while the high resolution spectrograph will be housed in a separate cryostat and fed with an SM fiber that passes through the SCALES cryostat.

\section{OPTO-MECHANICAL DESIGN}
\label{sec:optmechdesign}
SCALES uses an all-reflective optical design (with the exception of the refractive entrance window, coronagraph masks, lenslet array, dispersers, and filters).
As described in the previous SCALES/PSI-Red paper~\cite{Stelter2018}, the mirrors, mounts, and optical bench will all be machined from 6061 aluminum, which was chosen for both its machineability and its excellent mechanical properties. 
The mirrors will be machined from a single billet of RSA 6061 aluminum, which due to the manufacturing process used by RSP, has a much lower surface roughness after diamond-turning than traditionally manufactured 6061. 
The mirrors will be coated with gold, which has excellent reflectivity in the infrared.

The mounts, bench, and mirrors are all constructed from one material (6061 Aluminum) the Coeffcient of Thermal Expansion (CTE) is nearly identical between modules.  This allows the entire SCALES optical train to be aligned, tested, and adjusted at room temperature with no other athermalization considerations. 
While the lenslet array cannot be aligned warm, the depth of focus of the fore-optics is several millimeters; the depth of focus on the spectrograph side is much tighter (less than a millimeter) and is the limiting factor in the alignment of the lenslet array. 
In practice, we expect that proper alignment of the lenslet array to the spectrograph module will take an additional one or two cooldowns of the instrument during the integration and test phase.

The mirrors have reference features such as precision dowel pin holes and bolt pads designed into their substrates and are defined relative to the optical surface; these reference features allow alignment to become part of the design process and make up the `bolt-and-go' philosophy described in \S\ref{subsec:boltandgo}.
The opto-mechanical designs of the mirrors is described in \S\ref{subsec:mirrors}.
The lenslet array is discussed in \S\ref{subsec:lensletarray}, and the slenslit opto-mechanical design is explored in \S\ref{subsec:slenslitOMdesign}.
\S\ref{subsec:cryomechs} discusses the cryo-mechanisms interior to the cryostat.
The cryostat and its interface plates is discussed in \S\ref{subsec:cryostat}.

\subsection{Bolt-and-Go}
\label{subsec:boltandgo}
Modern Single-Point Diamond Turning (SPDT) uses a small diamond spindle whose position is precisely controlled while spinning at $\sim100 000$ RPM.
The part being turned is spun on a spindle plate giving a total of four to five axes of control depending on the machine set-up, allowing for effectively unlimited creativity with respect to surface figure~\cite{Yin2015,Carrigan2015}.
While off-axis aspherical mirrors are not trivial to machine or test, surface figure errors of $\frac{1}{10}\lambda$ for $\lambda=633$ nm and RMS surface roughness of $1$ nm are possible.
The specifications of SCALES, given our long operating wavelengths, are well-matched with SPDT, and our optical train is designed to be telecentric to allow for easier testing.
The spectrograph in particular, which has several high-order polynomial aspheres (more aspherical than OAPs/OAEs), has looser specifications because the wavefront has already been sampled by the lenslet array; as long as the geometric aberrations are kept reasonably small, the particular details of the surface figure or RMS roughness are unimportant.
Similarly the slenslit specifications benefit from the fact that we are not preserving the PSF; rather, we are preserving the diffraction-limited performance set by the lenslet array pupil images.
Thus, we do not have to have good optical performance across the entire width of each slice; rather, we are concerned only about the rather small pupil images.

The bolt-and-go approach has been used with great success for a variety of instruments, from the small (FISICA/FLAMINGOS-1\cite{fisica2004} to large facility-class instruments such as CIRCE~\cite{circe2013,circe14,2018JAIcirce}, FRIDA~\cite{FRIDA2013}, and MIRADAS~\cite{miradas2016}, all of which implement the bolt-and-go method to precisely locate many if not all of their diamond-turned mirrors.
As described in the previous SCALES/PSI-Red SPIE paper\cite{Stelter2018}, the bolt-and-go approach is a semi-kinematic mounting scheme that puts much of the onus of alignment into the opto-mechanical design instead of during the alignment phase.
The semi-kinematic scheme depends on using reference features such as pads centered on bolt holes to provide a plane of contact with the spindle plate (and eventual mirror mount or bench for the mirror and its mount, respectively); the pads' height is defined relative to the vertex of optically-powered mirrors, and the mirror surface for flat mirrors.
The back surface of the substrate is machined such that the pads protrude above the rest of the back surface.
The location on the plane of contact is defined by using two to three precision-ground stainless steel dowel pins.
The location of both pins are defined relative to the vertex of optically-powered mirrors and the aperture center of flat mirrors.
A major benefit of this system is that alignment is done largely in the design phase; the reference features (bolt pads and pin locations) make the alignment process largely the same as assembly (hence, `bolt and go').

\subsection{Mirror and Mirror Mount Opto-Mechanical Design}
\label{subsec:mirrors}
For mirrors, three bolt pads are designed to rest $0.5$ to $1.25$ mm above the back surface of the substrate; this provides a plane of contact with the mount (the idealized version would be three point contacts defining a plane).
Two pinholes make up the other elements of the bolt-and-go scheme; the pinhole at the mirror center locates the mirror substrate on the mount while the second pinhole defines the clocking angle of the mirror.
The bolts themselves are used only as a clamping force; they provide no location information for the optic.


\begin{figure}
    \centering
    \includegraphics[width=0.85\textwidth]{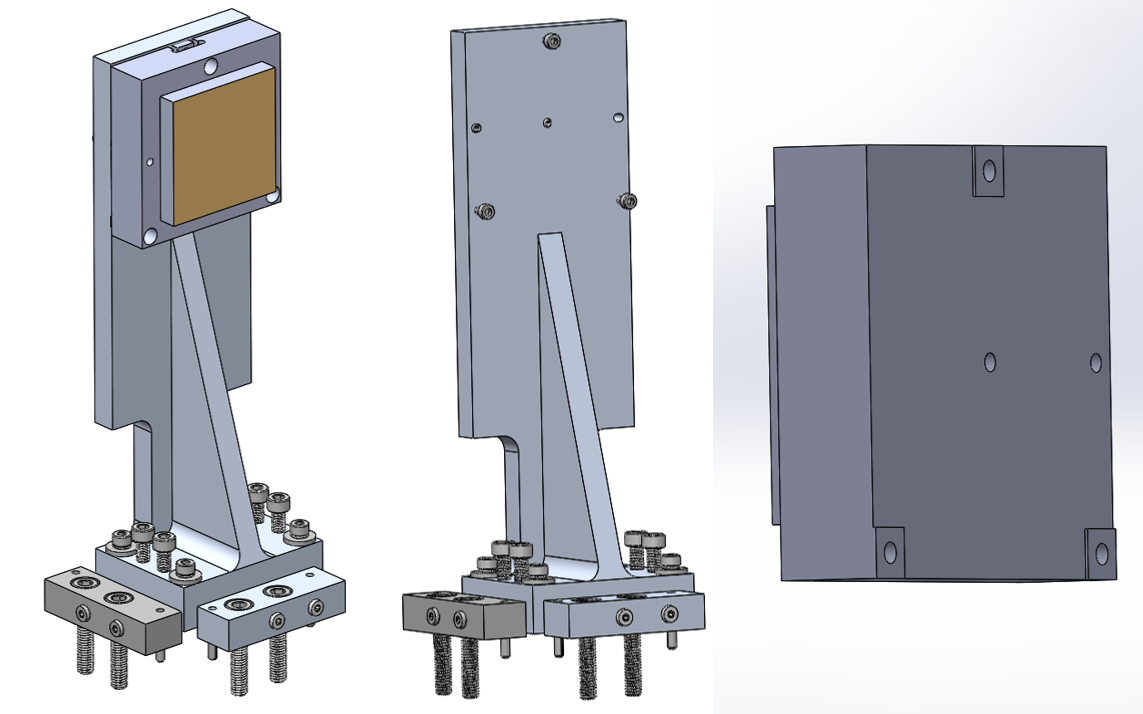}
    \caption{An example SCALES bolt-and-go mirror mount.
    Left: The mount is located using the nudger blocks (the separate blocks at the foot of the mount) and bolted into place using 3 bolts.
    Jackscrews allow for height and tip/tilt adjustment at the base of the mount.
    The mirror sits on a separate substrate and is gold-coated.
    Middle: The same but with the mirror hidden.
    Two pins, one at center and the other near the edge of the mirror substrate, define the location and clocking of the mirror.
    Right: The back of the mirror.
    Note the bolt pads are slightly higher with respect to the back surface.}
    \label{fig:mirrormount}
\end{figure}

Belleville washers will be used under the fasteners to insure consistent loading of the kinematic pads.
Adjusting the bolt pad height by shimming can adjust piston, tip, and tilt, while adjusting the clocking and x and y positions of the center of the mirror require adjusting the mount (either by shimming or shifting the mount).

For alignment of the optics on the bench a {\it nudger} will be used that consists of aluminum blocks bolted to the optical bench and located with two precision dowel pins with high thread count adjuster screws that can move the mount of an optic by small amounts about its local coordinate system.
The mounts use three bolt pads as the mirrors do, but do not depend on dowel pins for locating the mounts on the optical bench.
We have also added three jackscrews that are adjustable without entirely dismounting the mirror and its mount (in comparison to using shimming/lapping the bolt pads themselves in order to adjust the bolt pad heights).

\subsection{Lenslet Array}
\label{subsec:lensletarray}
The lenslet array is based on the ALES lenslet array, and is composed of three parts: the lenslet array, silicon picture frame, and diffraction spike suppressing pinholes.
The lenslets are  photo-lithographically etched into a silicon substrate.
The substrate will be glued to a `picture frame' of pure silicon, on the other side of which be glued a diffraction spike-suppressing pinhole grid (also photo-lithographically etched silicon).

\subsection{Slenslit Opto-Mechanical Design}
\label{subsec:slenslitOMdesign}
\begin{figure}
    \centering
    \includegraphics[width=0.9\textwidth]{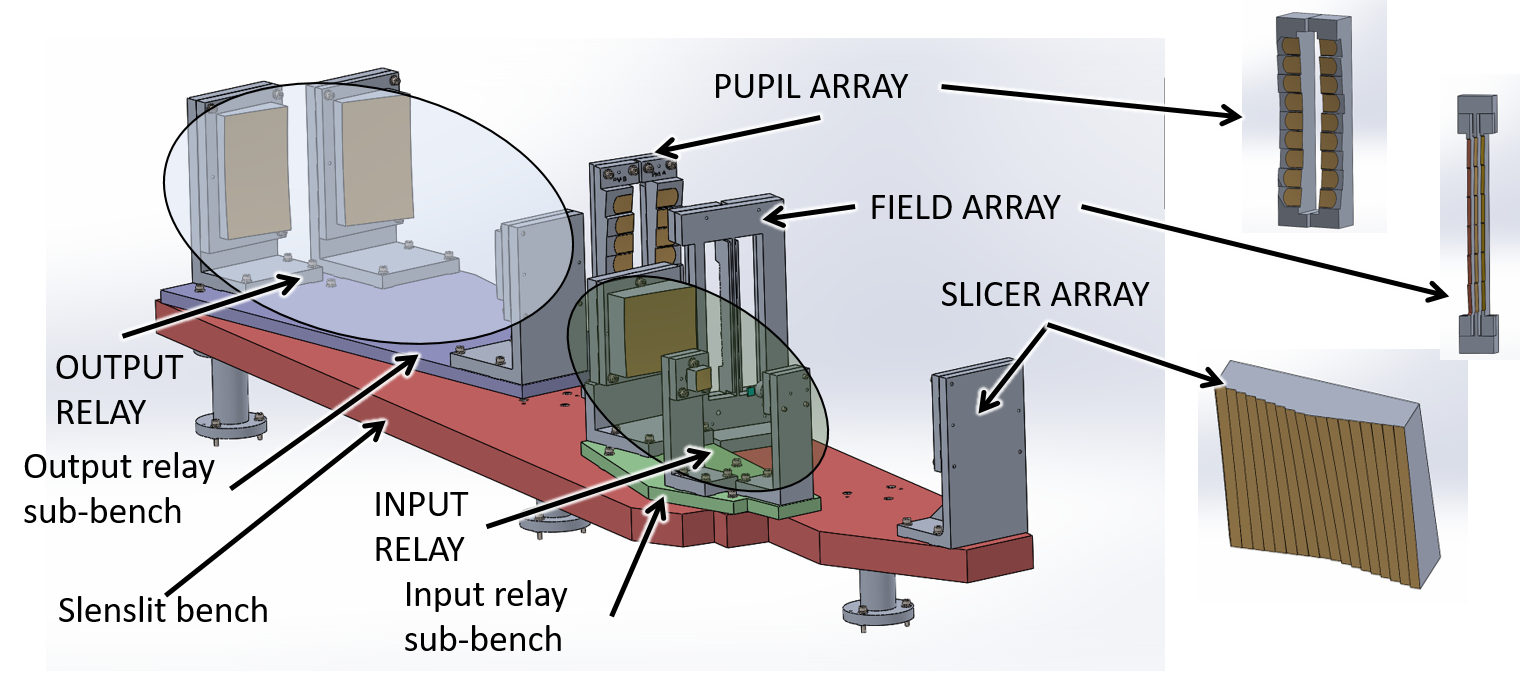}
    \caption{Anotated CAD rendering of the slenslit optics.
    The slicing mirror arrays are shown to the right.
    The orientation is roughly the same as in Figure~\ref{fig:slenslitODdesign}.
    For clarity, the lenslet array, return flat, and optical beams are not shown.
    }
    \label{fig:slenslitOMdesign}
\end{figure}

The slenslit optical design was done in Zemax initially in three modules (input relay, slicing optics, and output relay) before being integrated into the spectrograph optical model.
The opto-mechanical design of the slicing optics (shown in Figure~\ref{fig:slenslitOMdesign}) is designed in SolidWorks based on the optical prescription in Zemax, and uses realistic tool shapes and tool paths. 
Each mirror and mirror array uses the `bolt-and-go' design approach, with each mirror's mechanical reference features  (pin holes, bolt pads) are defined relative to the optical features (radius of curvature, center of curvature, vertex, etc.).
The mounts for each mirror are also `bolt-and-go.'
The slenslit assembly is checked against the optical beams exported from Zemax to verify that each mirror's location, power, and tip/tilt is correct, and to ensure that no vignetting is incurred.
Each module is telecentric, allowing for the separate test, assembly and alignment of each before integration onto the slenslit bench.
The mode selector mechanism (see \S\ref{subsubsec:modeSelectorSlide}) carries the return flat fold mirror and baffles.

\subsection{Cryo-Mechanism Design}
\label{subsec:cryomechs}
Here we describe the cryo-mechanism design, and we proceed in the order in which photons encounter them (with the upgrade mechanisms discussed after the IFU mechanisms).
We are currently in the Preliminary Design phase, although several mechanisms have been completed through Detailed Design and are being fabricated as a risk-reduction measure.
The Coronagraph Slide and Lyot Wheel mechanism in particular have been pushed to a higher design level than the remaining mechanisms.
The Self-Coherent Camera (\S\ref{subsec:scc}) and HCHD fiber coupler \& spectrograph (\S\ref{subsec:hchd}) designs have not been integrated into the SCALES CAD model.  These modules will be based on previous instrumentation~\cite{Gerard2018,Pezzato2019,hispec2019}.

All of the mechanisms uses Phytron high-vacuum stepper motors, which are micro-stepped to give higher step count precision; we have also baselined using only one size motor (Phytron VSS 43) as to reduce the number of types that require swappable spares to be kept at Keck once deployed.
We control the motors using Galil controllers (model number DMC-4040(16BIT,ISCNTL)-CO22(5V)-I200-D4140).
Absolute encoding is rather difficult and expensive to implement in cryogenic environments with sensitive IR detectors, so we use step counting to inform us of the location of mechanism positions, with reference points set using Honeywell sub-miniature single pole double throw switches (model 11SX1-T) wired Normally Closed.\footnote{
Wiring the switches Normally Closed helps to troubleshoot bad electrical connections -- if a switch breaks or the switch circuit otherwise opens, we can troubleshoot cable segments to find the fault.}
In general, linear mechanisms use one switch as a datum or origin point, with other locations determined primarily by step counting and may or may not be indicated with individual switches; two switches indicate the `soft' (software-defined) and `hard' (mechanically-defined) limits to the range of motion.
If the moving stage trips the hard limit switch, power to the motor is cut immediately to prevent the mechanism from damaging itself.

Rotary mechanisms use one switch position as a `home' position, with each other position defined first by step counts relative to the home position, and where high precision is needed (such as the Lyot wheel) a passive detent mechanism is used.
If desired, each position of a rotary mechanism can trip a limit switch using a roller bearing mounted to the wheel with a shoulder screw; the home position uses two roller bearings that trip two limit switches stacked on top of one another.
Generally there is no need to set a hard limit for rotary motion.

In order to remove backlash, the mechanisms are designed to always approach the desired position from the same direction.
This means that rotary mechanisms spin in one direction only, while linear mechanisms incorporate a backlash-removing motion before moving toward the desired position.

\subsubsection{Coronagraph slide}
\label{subsubsec:coronagraphSlide}
\begin{figure}[hp]
    \centering
    \includegraphics[width=0.9\textwidth]{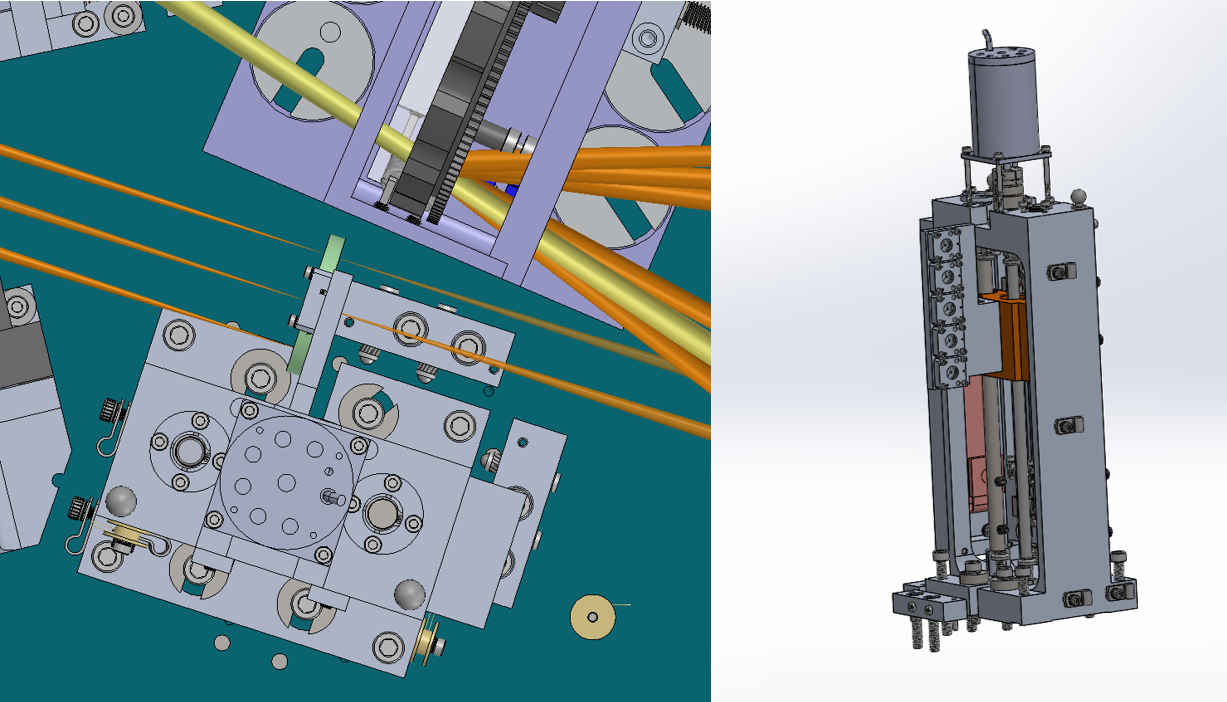}
    \caption{Left: Overhead view of the coronagraph slide on the optical bench.
    The orange and yellow beams are the optical beams of the imager and IFU, respectively, imported from Zemax.
    The second OAP is to the left of the slide, with the Lyot mechanism located above.
    The round disk to the right of the slide is a Lakeshore temperature sensor.
    Right: Isometric view of the coronagraph slide.}
    \label{fig:coronagraphslide1}
\end{figure}

The coronagraph slide is a linear motion mechanism that carries up to 4 coronagraph masks and a pupil imaging lens and is shown in Figure~\ref{fig:coronagraphslide1}.
The design of the imaging arm requires that the slide be able to fully retract and not vignette the 20~arcsec field of view, which requires a comparably large range of travel.
The design  design heritage from linear mechanisms deployed in CIRCE\cite{2018JAIcirce} and MIRADAS\cite{miradas2016} by the University of Florida at the 10 m Gran Telescopio Canarias.
The slide (colored orange in Figure~\ref{fig:coronagraphslide1}) rides along two steel rails on rulon linear sleeve bearings; an uncoated stainless steel ball screw is coupled using a helical coupler to a Phytron VSS43 stepper motor (the coupler allows for small shaft misalignments between the motor and lead screw).
The position of the slide is indicated (but not defined) by 9 limit switches bolted to the back of the frame.
Ramps machined into the back of the slide trigger the limit switches during travel and provide rough location information of the slide.
As the mask holder is, by definition, at a focal plane, we can easily measure the location of the slide to verify its position by taking an image either with the IFU or imaging detector.

The mask holder is bolted to the slide and is meant to be easily swapped out in order to facilitate new mask installation and adjustment.
The slide is thermally coupled to the frame with a flexible copper strap; this is expected to dramatically decrease the time required to cool the slide and mask holder, which otherwise have a very long conduction path through bearings and shafts to the frame.
Like the mirror mounts, the location and orientation of the coronagraph slide is defined and adjusted by nudger blocks. 
The bolts used to secure the slide to the optical bench merely hold the mechanism in place and do not define the location.
Jackscrews at the corners of the frame are used to adjust the tip and tilt of the mechanism.

\subsubsection{Lyot wheel}
\label{subsubsec:lyotWheel}
\begin{figure}
    \centering
    \includegraphics[width=0.9\textwidth]{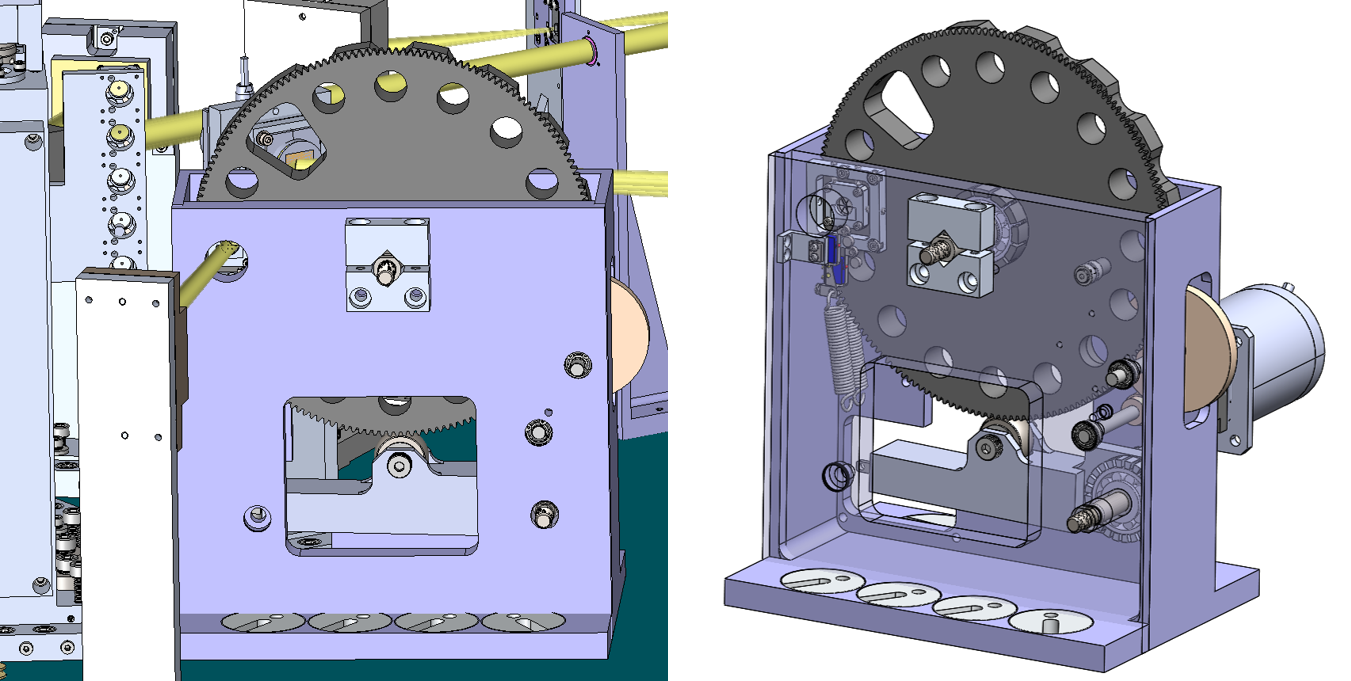}
    \caption{Right: Lyot mechanism shown on the bench. 
    The yellow beams are the IFU beams exported from Zemax; the imager components and beams are hidden from view.
    The detent mechanism is the aluminum bar with the wheel located directly underneath the main wheel (the springs are shown in their relaxed position to the left and below of the beam).
    Left: The Lyot mechanism in an isometric view.
    The front plate of the Lyot housing has been made transparent.
    }
\label{fig:lyot1}
\end{figure}

The Lyot wheel is designed to hold up to 15 coronagraphic pupil masks (as well as one fold mirror for the imaging mode) as shown in Figures~\ref{fig:lyot1}.
Because the pupil mask alignment is critical and has tight specifications ($21 \mu$m, or 0.3\% of the pupil diameter), we elected to use a passive detent mechanism that consists of a follower wheel spring-loaded to fall into a V-shaped groove at every mask position.
The detent mechanism holds the wheel in place and is expected to be repeatable to a few microns without exceeding the torque required to leave the detent groove beyond the motor specifications or extreme gearing to get the step resolution required for the mask location specification.
The masks are designed to be held by a mask whose position can be adjusted with a removable picture frame that has fine adjustment screws; this allows us to co-locate the masks with their detent position.

\subsubsection{Mode selector slide}
\label{subsubsec:modeSelectorSlide}
\begin{figure}
    \centering
    \includegraphics[width=0.9\textwidth]{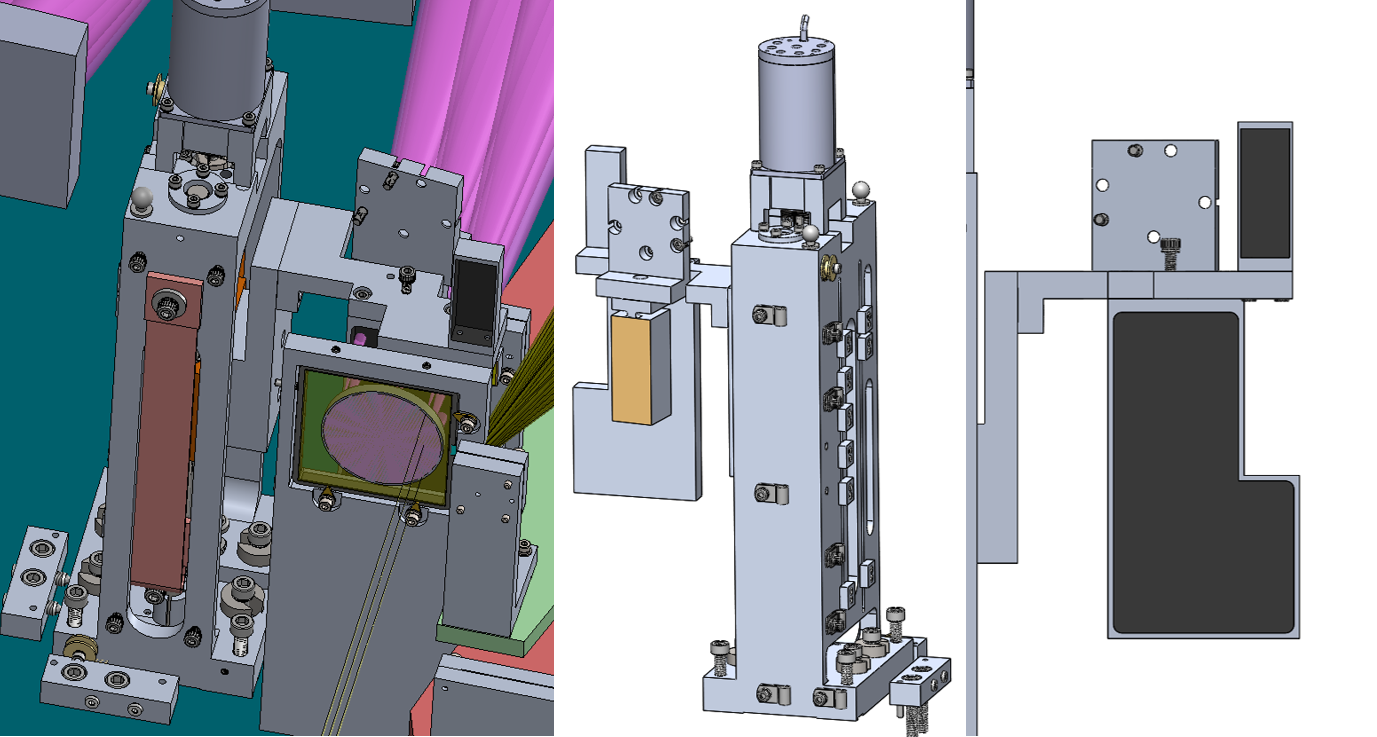}
    \caption{Left: The Mode Selector Slide mechanism on the bench in the mid-resolution mode (with the low-resolution beams blocked).
    The lenslet array is shown in its mount (the yellow rectangle is the lenslet array).
    The pink circle is a dummy surface (part of the beams exported from Zemax) and is used to align the beams to the CAD model.
    The magenta beams are the low-resolution IFU spectrograph beams imported from Zemax.
    The yellow-green beams to the right of the lenslet array is the slenslit input relay beams imported from Zemax.
    Middle: The view from the spectrograph module of the Mode Selector Slide.
    The return mirror sits on an adjustable mount which is mounted to the diving board that is bolted to the slide.
    Right: front view of the Mode Selector Slide baffles.
    From the bottom up, we have a `dark' mode, medium-resolution mode, and low-resolution mode.
    The baffles are colored black and will consist of Acktar black absorbing foil glued to an aluminum holding cell.}
    \label{fig:ModeSelector1}
\end{figure}

The mode selector slide is used to switch between the low-resolution and medium-resolution spectroscopy modes. 
It is very similar to the coronagraph linear slide, with the mask holder replaced with baffles and an adjustable return mirror that are mounted to a `diving board' as shown in Figures~\ref{fig:ModeSelector1}.
The baffles and return mirror work to either allow light from the low-resolution footprint of the lenslet array to pass directly to the spectrograph or to allow the slenslit lenslets to illuminate the slenslit optics before passing to the spectrograph.
The return mirror is mounted with an adjustable base for alignment.
The baffles will use an Acktar black absorbing foil to facilitate light absorption.
We also have designed one position to act as a `dark' to help check for scattered light within the spectrograph module.

\subsubsection{Disperser carousel}
\label{subsubsec:disperserCarousel}
The Disperser Carousel mechanism (see Figure~\ref{fig:ifumechs}) is designed to hold optics at the pupil plane created by the spectrograph module.
The low-resolution mode uses LiF prisms, while the mid-resolution mode uses gratings.
The disperser carousel will also hold a mirror used for alignment of the mechanism as well as an engineered diffuser that will allow us to flatfield the detector while simultaneously relaxing the requirements imposed on the precision of this mechanism.
By diffusing the light of each lenslet pupil over a large patch of the detector (about 1 mm in size) we will be able to calibrate the detector response and increase the precision repeatability requirement that the spaxels be placed within 0.1 pix to 10 pix.

The prisms are designed with an AR coating on the first surface and a gold coating on the second surface.
Originally we had hoped to have the first surface coating also function as a bandpass filter, but given the rejected out of band light would be reflected into the spectrograph optical train toward the detector, we have elected to place the filters in a separate mechanism.

\subsubsection{IFU filter wheel}
\label{subsubsec:ifuFilterWheel}
\begin{figure}
    \centering
    \includegraphics[width=0.9\textwidth]{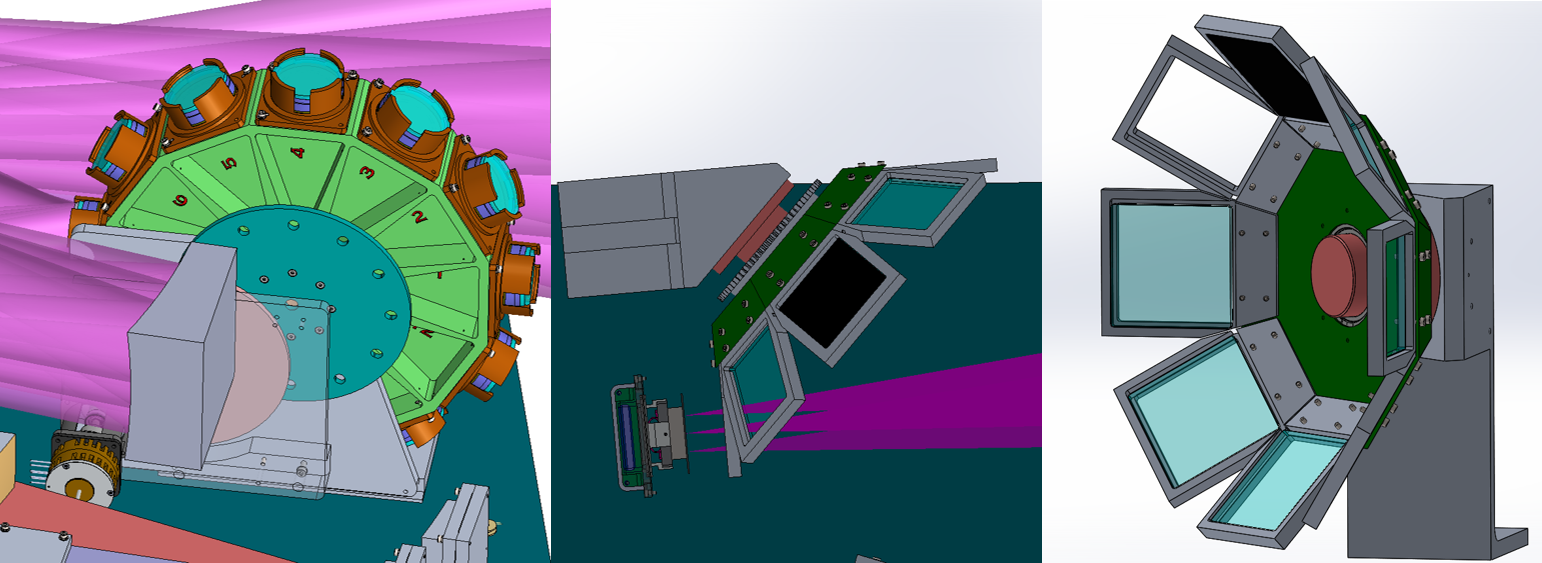}
    \caption{The spectrograph cryo-mechanisms.
    Left: Disperser Carousel mechanism shown on the bench.
    It has 12 positions, and in this rendering they are populated with identical models of prisms.
    Middle: An overhead view of the mechanism on the bench (blue-green).
    The H2RG detector is to the lower left. 
    The spectrograph optical beams, exported from Zemax, are shown in magenta.
    The frames hold the filters.
    Right: The IFU filter as seen from the side.
    It holds up to 6 filters plus an open position (empty frame) and dark position (black square).}
    \label{fig:ifumechs}
\end{figure}

The IFU filter wheel is designed as a bent-wheel or flower-petal mechanism to reduce its footprint, and is shown in Figure~\ref{fig:ifumechs}.
The filters are designed to match a particular prism or grating, thereby allowing us to offload the design effort of the AR/bandpass function of the prism coating into two separate components.

\subsubsection{Imager upgrade:  cold stop and filter wheel}
\label{subsubsec:imagerColdStop}
\label{subsubsec:imagerFilterWheel}
The imager upgrade benefits immensely from having a selectable cold stop as the IFU uses the K mirror mechanism on the AO bench to keep the telescope secondary pupil orientation fixed, which by necessity means that the field rotates relative to the detector.
The `vanilla' imaging mode benefits from the K mirror mechanism tracking the field rotation, which means that the cold stop design must either rotate to match the secondary rotation or be circularly symmetric.
Given that our philosophy has been to minimize any potential harm to the main IFU mode, we have elected to put in a cold stop mechanism that holds several varieties of cold stops.
This is a rotary mechanism that will be based on the Lyot wheel mechanism, as the position and repeatability requirements for the cold stop are similar, and both lie at a pupil plane.

The imager upgrade also requires imaging filters, which are not necessarily the same as the filters used for the IFU. 
We worked with Keck Observatory staff to identify the filters most commonly used with NIRC2, which has similar (albeit not identical) bandpass, field of view, and diffraction-limited performance requirements.
18 filters were identified as high-use, and so we have designed a filter mechanism (based off of a simplified Lyot mechanism) with two filter wheels that can accommodate the 1 inch standard filters.
The location of the filter wheels is directly behind the cold stop mechanism, as the filters are best suited to a collimated airspace (the space between the first OAP and the cold stop is ill-suited to hosting mechanisms of any kind due to the narrow angle between incoming and outgoing beams from the OAP).
The filter wheels do not require a detent mechanism (the filters are $\sim2$x the beam diameter at their location) and so do not have tight repeatability requirements placed on them.
It is important to note that the filters cannot be used in conjunction with the IFU mode at the location of this mechanism; the filters, being thin-film interference filters, fail our strict wavefront RMS specifications derived from the high contrast requirements and thus must follow, not precede, the coronagraphic masks.
However, the imaging arm is split off after the pupil plane coronagraph masks at the Lyot wheel, and so our design requires independent filter mechanisms for the imager and spectrograph modes.

\subsection{Cryostat Design}
\label{subsec:cryostat}
\begin{figure}
    \centering
    \includegraphics[width=0.95\textwidth]{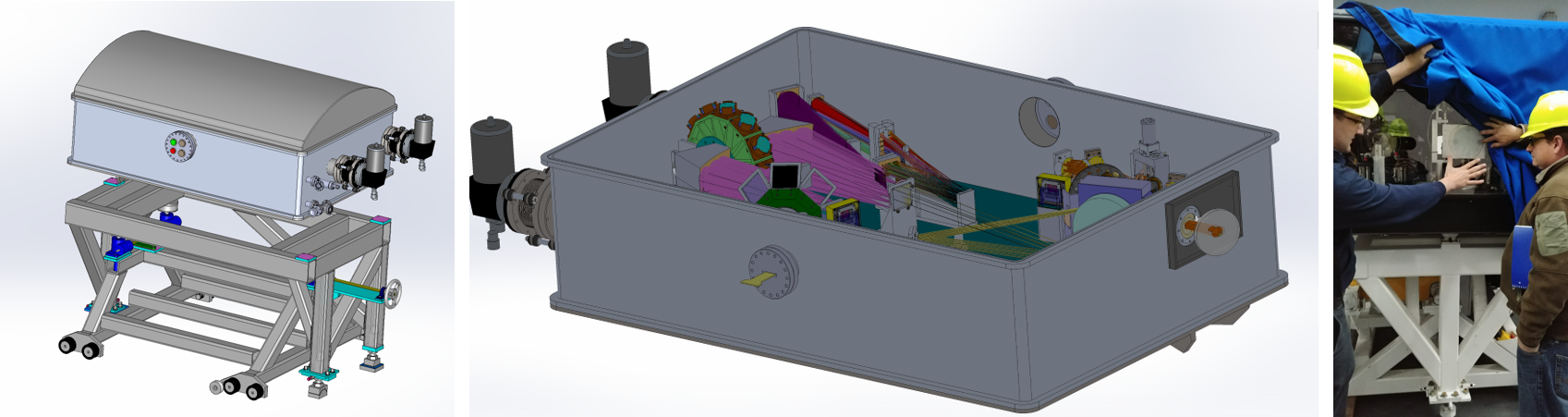}
    \caption{The SCALES cryostat and cart.
    Left: Isometric view of SCALES resting on the DSM cart (not shown are jackscrews used for coarse alignment to the AO optical axis).
    The conflat stubs and flats are shown with Amphenol and Teledyne flex cables populating the flats.
    Middle: The cryostat with the lid and radiation shields hidden.
    The view is rotated 180\textdegree. 
    The black rectangle is the entrance window environmental cover, and the disc in front of it is part of the AO optical bench imported from Zemax.
    Right: The DSM cart at Keck Observatory being inspected by A. Skemer and N. MacDonald (photo courtesy D. Stelter).
    }
 \label{fig:cryostat}
\end{figure}

SCALES is designed to operate at cryogenic temperatures to lower the emissivity of the optics, which requires placing the optics in a cryostat operating at $<10^{-6}$ mTorr.
The aluminum vacuum vessel will be large enough to contain the optics, mechanisms, and detectors on the $\sim 1.1$ m $\times 1.9$ m optical bench, as well as two layers of radiation shielding made of aluminum sheet and several layers of aluminum-coated Mylar superinsulation.
The current design is shown in Figure~\ref{fig:cryostat} and is split into three main components: the bottom plate, which interfaces with the instrument cart via adjustable jackscrews; the middle, which carries the vacuum interfaces; and the top, which is removed during servicing and has no vacuum penetrations.
The cryostat will rest on the cart previously used for the Dual Star Module (DSM) of the decommissioned Keck Interferometer~\cite{keckInterferometer2008}.

We will cool SCALES with two CTI Model 1050 Cryodyne refrigerators (black blocks and grey cylinders on the back of the cryostat in Figure~\ref{fig:cryostat}), each with 65 W of heat lift at 77K and 7 W at 20K simultaneously. 
The cold heads will sit on stainless steel bellows to dampen vibration imparted to the instrument.
The vacuum feedthrus will use commercial-off-the-shelf (COTS) Conflat stubs welded to the sides of the cryostat, with modified Conflat flanges holding various interfaces such as the vacuum valve, pressure monitor, and Teledyne flex cables.
Electrical connections will use Amphenol hermetic connectors soldered to fan-out boards on the inner side of the flanges.

\section{SUMMARY}
\label{sec:summary}
We have presented an update on the status the design of SCALES, a purpose-built high-contrast IFU operating from $2-5 \mu$m at R$\sim100-10000$.
Particular focus has been given to the main science drivers of exoplanets, although diverse secondary science drivers, which range from Solar System science to extragalactic SMBH accretion disk imaging, are also explored.
The optical and opto-mechanical design of the SCALES optics and cryo-mechanisms, particularly the slenslit, are discussed in detail.
SCALES is midway through the Preliminary Design phase and is slated to be on-sky in mid-2025.

\acknowledgments 
We gratefully acknowledge the support of the Heising-Simons Foundation through grant \#2019-1697.
UC Observatory engineers and staff have worked tirelessly through the COVID-19 pandemic and raging fires.
The lead author wishes to acknowledge the Herculean effort on the part of frontline healthcare workers through the pandemic including his father, who serves as an ambulance driver.

\bibliographystyle{spiebib} 
\bibliography{scales.bib} 

\end{document}